\newcommand{\PaperTitle}{Global BGP Attacks that Evade Route Monitoring}
\newcommand{\PaperNumber}{221}

\documentclass[10pt,sigconf,letterpaper,nonacm]{acmart}

\usepackage{color}
\usepackage{graphicx}
\usepackage[labelformat=simple]{subcaption}
\usepackage{xspace}
\usepackage{multirow}
\usepackage[ruled,vlined]{algorithm2e}
\usepackage{ulem}
\newcommand{\myitem}[1]{\vspace*{0.02in}\noindent\textbf{#1}}

\def\showcomments{0}

\ifnum\showcomments=1
    \newcommand{\henry}[1]{{\textcolor{green}{[Henry: #1]}}}
    \newcommand{\jen}[1]{{\textcolor{blue}{[Jen: #1]}}}
\else
    \newcommand{\henry}[1]{}
    \newcommand{\jen}[1]{}
\fi

\begin{document}
\title{\PaperTitle}

\author{Henry Birge-Lee}
\orcid{0000-0001-7141-345X}
\affiliation{
\institution{Princeton University}
  \city{Princeton}
 \country{United States}
}
 
\email{birgelee@princeton.edu}
\author{Maria Apostolaki}
\orcid{0000-0003-0342-2631}
\affiliation{
\institution{Princeton University}
  \city{Princeton}
 \country{United States}
}
\email{apostolaki@princeton.edu}
\author{Jennifer Rexford}
\orcid{0000-0002-0231-8165}
\affiliation{
\institution{Princeton University}
  \city{Princeton}
 \country{United States}
}
\email{jrex@cs.princeton.edu}

\begin{abstract}
As the deployment of comprehensive Border Gateway Protocol (BGP) security measures is still in progress, BGP monitoring continues to play a critical role in protecting the Internet from routing attacks. Fundamentally, monitoring involves observing BGP feeds to detect suspicious announcements and taking defensive action. However, BGP monitoring relies on seeing the malicious BGP announcement in the first place! In this paper, we develop a novel attack that can hide itself from all state-of-the-art BGP monitoring systems we tested while affecting the entire Internet. The attack involves launching a sub-prefix hijack with the RFC-specified NO\_EXPORT community attached to prevent networks with the malicious route installed from sending the route to BGP monitoring systems. We study the viability of this attack at four tier-1 networks and find all networks we studied were vulnerable to the attack. Finally, we propose a mitigation  that significantly improves the robustness of the BGP monitoring ecosystem. Our paper aims to raise awareness of this issue and offer guidance to providers to protect against such attacks.
\end{abstract}

\maketitle

\section{Introduction}

The Border Gateway Protocol (or BGP) allows for the exchange of routes between the independently-operated networks (known as Autonomous Systems or ASes) that comprise the Internet. BGP announcements contain reachability information for IP prefixes and are propagated from one network to the next to build the global IP routing table. However, BGP was designed with no form of message authentication, which allows adversaries to construct bogus BGP announcements for IP prefixes they do not control or legitimately route to. These incidents are known as BGP attacks or BGP hijacks. BGP attacks cause Internet traffic destine to a victim AS to be maliciously routed to the attacker's infrastructure. BGP attacks can cause significant damage, particularly as these routing attacks can be used to target higher-level applications run on top of the Internet~\cite{birgelee2022klayswap,coinbase2022celer_bridge,birgelee2018pki,sun2015raptor,apostolaki2017bitcoin,sun_cacm21}.

While securing BGP is fundamental to the security of many critical applications that run on top of the Internet, robust BGP security is not practical today with available technologies. Relatively high-security proposals like BGPsec require updates to existing hardware and have not yet seen any production deployment~\cite{bgpsec_reality}. Alternatives like Resource Public Key Infrastructure (RPKI) do have significant deployment, but these weaker solutions still permit many types of attacks~\cite{gilad2017max_length,dfoh}.

A pragmatic and commonly used alternative to comprehensive BGP security is BGP monitoring. BGP monitoring involves observing BGP updates and flagging updates that appear suspicious. One advantage of BGP monitoring is that it leaves the BGP announcement syntax untouched. However, BGP monitoring has a core limitation: it cannot detect a route it cannot see~\cite{milolidakis2021smart_hijacks}. Existing BGP monitoring systems have overcome this limitation by combining data from many vantage points spread across the Internet. With two of the most popular monitoring services (RIPE RIS and RouteViews) each having over 1000 different BGP peering sessions~\cite{ris_peers,routeviews_peers} with various top networks, this problem appears to be solved. Previous work on attacks that evade BGP monitoring only succeeded in affecting 2\% of Internet traffic at best~\cite{milolidakis2021smart_hijacks,birgelee2019sico}.

In this paper we demonstrate that even with the vast number of peering sessions with monitoring services, an adversary can easily launch an invisible BGP attack (not seen by BGP monitoring) that still succeeds.
We introduce a novel attack on BGP monitoring that allows an adversary to stop networks peered with monitoring services from exporting the adversary's malicious announcements to monitoring services while still directing traffic via the adversary's route.
Our attack involves the adversary tagging its BGP announcement with a specific value (the RFC-specified NO\_EXPORT BGP community) so that networks using the adversary's route do \emph{not} send it to BGP monitoring services. 
Even though this misuse of the NO\_EXPORT community limits the spread of the adversary's route, the adversary can still install this route into networks commonly used to forward traffic to the victim using a \emph{sub-prefix} BGP attack. This makes the effect of this attack substantial or even global.

We ethically demonstrate this attack in the wild (i.e., attacking an IP prefix we control) and found it to be highly effective, capturing traffic from \textbf{all} of a sample of 1k random Internet hosts. Even while affecting nearly the entire Internet, this attack was invisible to all BGP monitoring services we studied (which included RouteViews, RIPE RIS, Cisco Crosswork Cloud, and ThousandEyes). We studied the behavior of the NO\_EXPORT community at four tier-1 networks and found all networks we studied hid announcements from all monitoring services when the community was attached. In fact, assuming an adversary only installed its malicious route at networks we confirmed were susceptible to this attack, the adversary could hijack traffic from 23\% percent of the Internet to a random destination according to our simulation. Finally, we conclude with recommendations that can significantly reduce the viability of this attack with simple configuration changes.

\section{Threat Model}

We assume an adversary that is capable of making a malicious BGP announcement with the aim of redirecting traffic. This traffic redirection can be used to attack several different critical applications that  run atop the Internet~\cite{apostolaki2017bitcoin,birge2018bamboozling,sun2015raptor}. The adversary also aims to avoid detection by BGP monitoring (i.e., launch a stealthy BGP attack) to prevent the victim from taking defensive action. We assume the adversary's announcement is not stopped by IP prefix filters (as is the case with observed real-world attacks~\cite{birgelee2022klayswap,coinbase2022celer_bridge,ars20183ve}).

Furthermore, we assume the adversary will willingly seek new BGP transit or peer sessions to enable its attack.
Internet eXchange Points (IXPs) have many networks colocated in a single location. This allows an adversary to cheaply establish many peer and transit sessions from a single Point of Presence. Furthermore, Virtual Machines (VMs) at IXPs can be rented quite cheaply (e.g., $\sim$\$30 a month~\cite{fog_ch_ixp}) and have ports on the IXP's peering LAN. Thus, these VMs can establish BGP sessions with other networks at the IXP~\cite{bandwidth_tech_ams,fog_ch_ixp, amxis_vm}. 

Using an IXP VM, an adversary can approach colocated ISPs and request BGP sessions to propagate its malicious announcement. Social-engineering or manipulation of Internet Routing Registry (IRR) data (as done in past attacks~\cite{ars20183ve}) can persuade ISPs into white-listing a victim's IP prefixes on an adversary's BGP session. Many major BGP attacks have been launched by direct customers of tier-1 ISPs~\cite{bitcanal_nanog_discussion,celer_hijack_nanog_discussion,klayswap_nanog_discussion}, and some adversaries have used multiple tier-1 ISPs to spread their attack~\cite{bitcanal_nanog_discussion}.

\section{Attack Details}

\begin{figure*}
    \centering
    \begin{subfigure}{0.33\textwidth}
    \centering
        \includegraphics[scale=0.26]{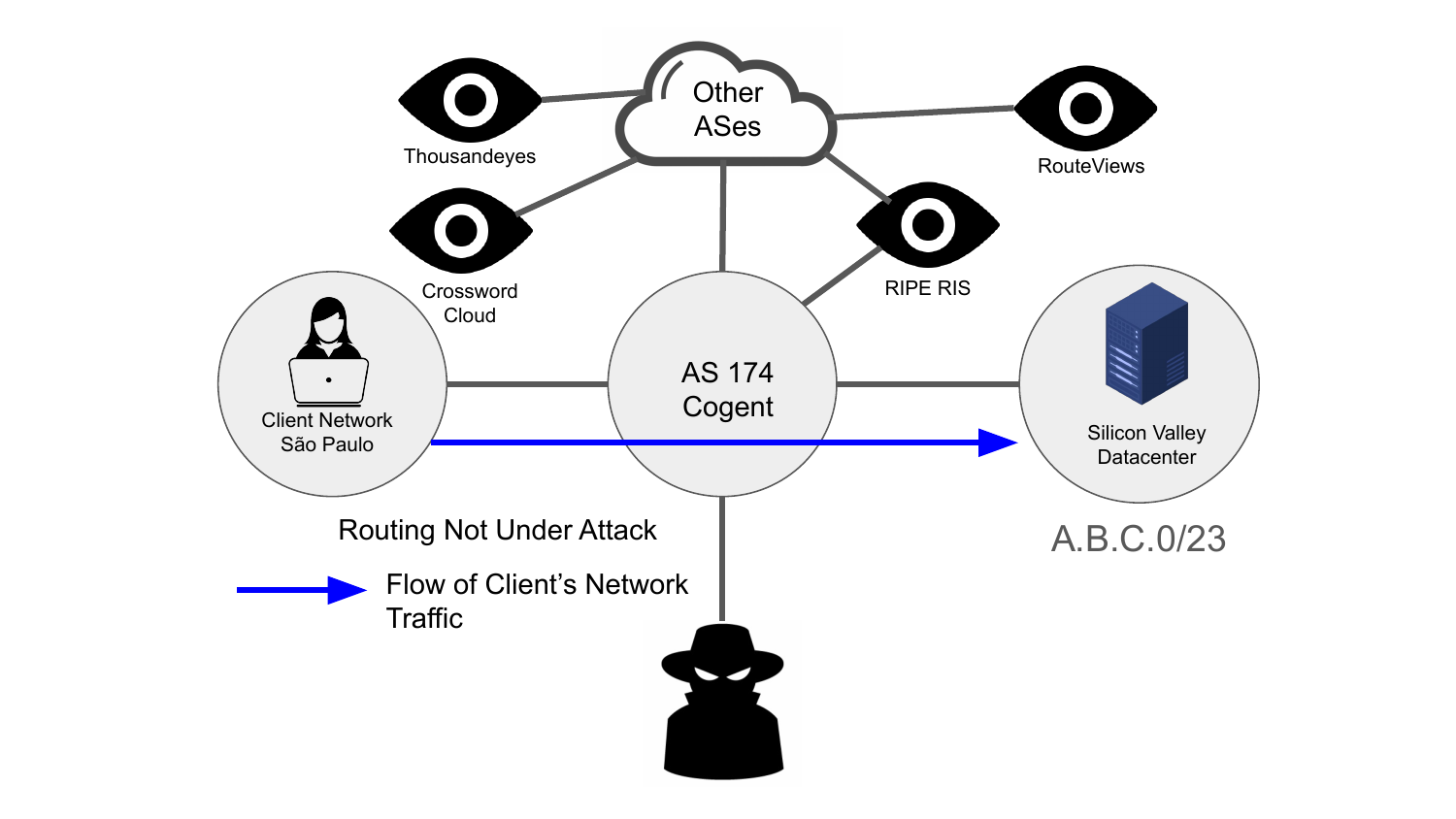}
        \caption{}
        \label{subfig:benign}
    \end{subfigure}%
    \begin{subfigure}{0.33\textwidth}
    \hfill
        \includegraphics[scale=0.26]{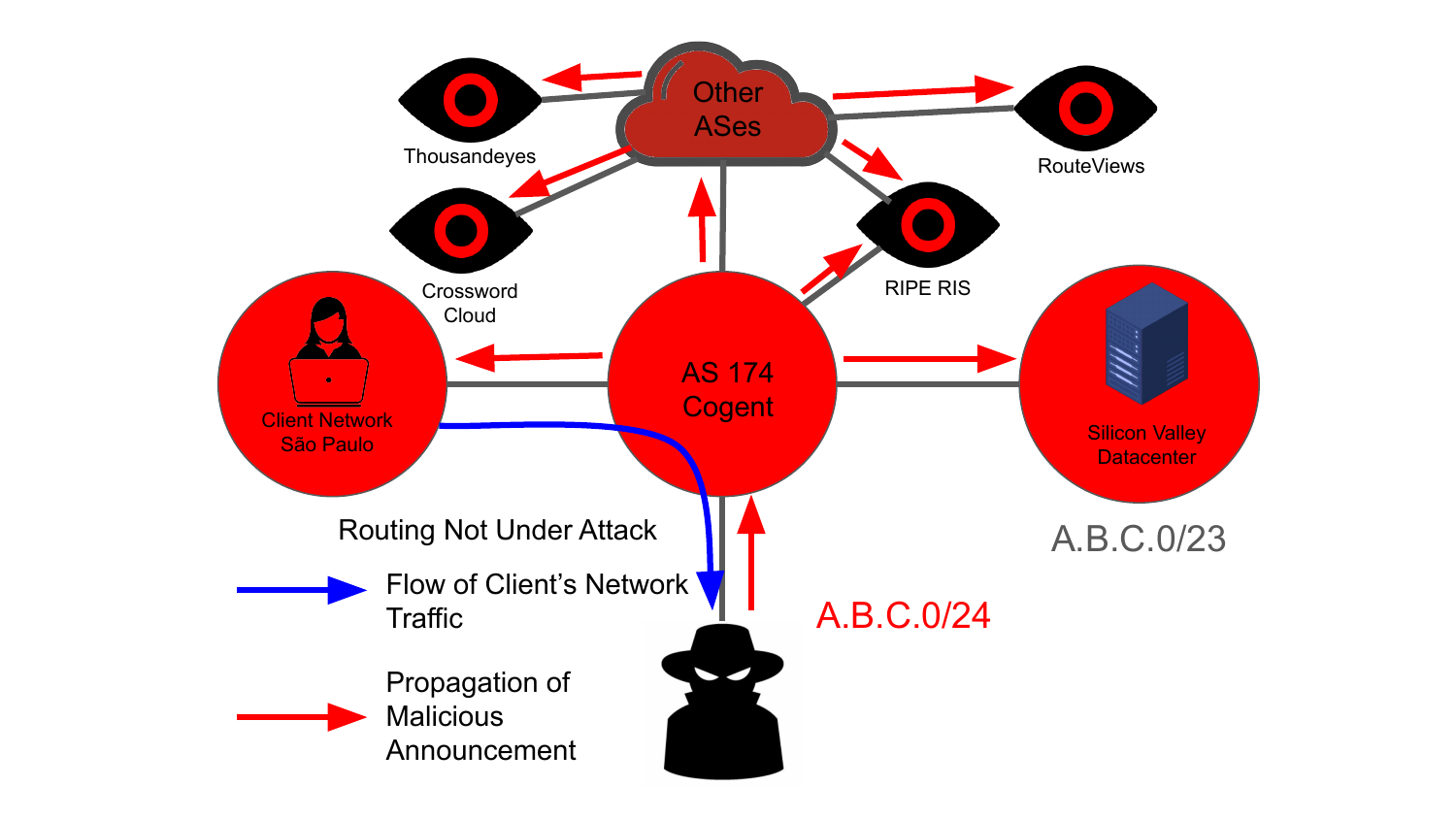}
        \caption{}
        \label{subfig:control}
    \end{subfigure}
    \hfill
    \begin{subfigure}{0.33\textwidth}
    \centering
        \includegraphics[scale=0.26]{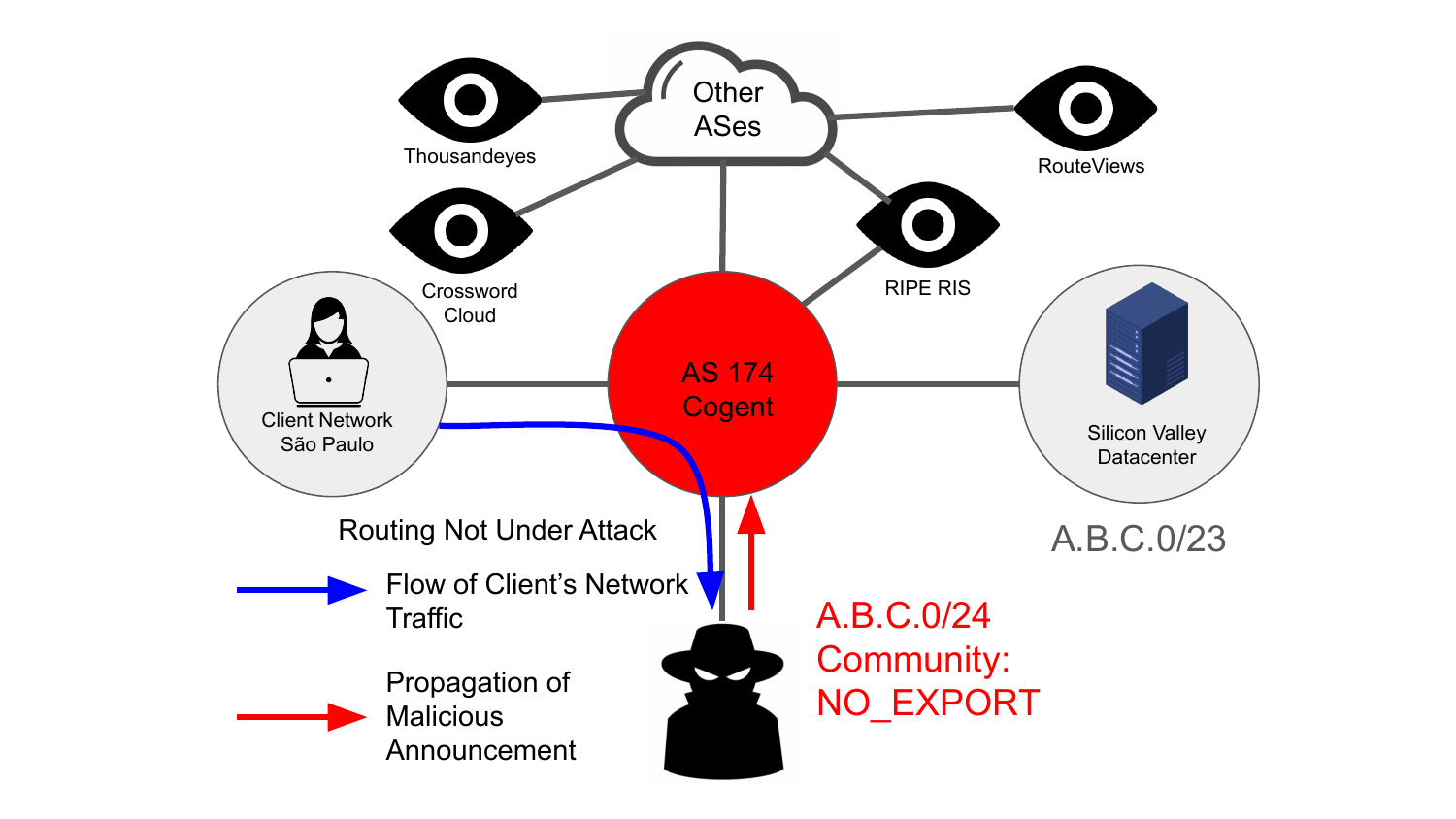}
        \caption{}
        \label{subfig:stealth}
    \end{subfigure}%
    \caption{ Routing during benign operations (\ref{subfig:benign}), a control non-stealthy attack (\ref{subfig:control}), and a stealthy attack (\ref{subfig:stealth}). Todo: fix figures}
    \label{fig:diagram}
    \vspace{-4mm}
\end{figure*}

Our attack relies on two key insights. First, the behavior of the RFC-defined NO\_EXPORT community  affects a network's sessions with BGP monitors. Second, because of longest-prefix-match forwarding, a network may be directing traffic over a route without hearing it or installing it into its RIB.

\subsection{Hiding the Route: NO\_EXPORT}

At the core of our attack is the ability to hide a route from BGP monitoring systems even when that route is installed in routers that provide a  feed to BGP monitoring systems. We accomplish this using an RFC-defined ``well-known'' BGP community called NO\_EXPORT.\footnote{The community NO\_EXPORT\_SUBCONFED has a similar behavior and can largely be used interchangeably with the NO\_EXPORT community for this purpose.} BGP communities are 32-bit tags that can be attached to BGP announcements and can impact how routers hearing those announcements behave~\cite{bgp-comm-support,krenc2023community_intent,RFC1997}. When attached to a BGP announcement, the NO\_EXPORT community instructs routers not to export that BGP announcement beyond a confederation boundary. Most routing software honors this behavior by default~\cite{no_export_example_cisco,no_export_example_juniper}, as support for the NO\_EXPORT community is stipulated by the same RFC that defines BGP communities~\cite{RFC1997}.
We utilize the key insight that \textbf{when the NO\_EXPORT community is attached to a route, routers do not export that route on sessions with BGP monitoring systems}.

This is because most BGP monitoring systems use multi-hop External BGP (eBGP) sessions with participating peer networks to receive BGP updates~\cite{routeviews_peering, thousandeyes_private_bgp,ris_peering_policy}. To the routers that handle these sessions, they appear to be with an external neighboring network, and thus, the export restrictions of the NO\_EXPORT community are applied. This creates the conditions that enable our attack: a router can have a route installed but not send it to monitoring systems.

\subsection{Attracting Traffic: Sub-prefix Hijacks}

Hiding the route from BGP monitors does not, by itself, allow for a  widespread attack. The use of the NO\_EXPORT community means the adversary can only affect as many networks as it has direct BGP sessions. This is because all ASes use the same RFC-defined value for the NO\_EXPORT community, so the first AS that sees an adversary's announcement will act upon the community and not export the adversary's route any further. This makes it  difficult to scale to thousands of ASes. %

However, the adversary can attract more traffic using a \emph{sub-prefix} attack where it announces a longer, more-specific prefix than the victim (i.e., true prefix owner) it is targeting. Longest-prefix-match IP forwarding means that routers will always use the most-specific (longest) IP prefix possible that matches a particular packet. Thus, in any network with the adversary's sub-prefix attack installed, the adversary's route will take priority over the victim's. With this in mind, \textbf{the adversary can install its malicious sub-prefix route in large transit providers or upstream of the victim} that, due to their topological location, would naturally carry a large amount of the victim's Internet traffic. Even though the propagation of the adversary's announcement is limited by the NO\_EXPORT community, the victim's benign BGP announcement attracts traffic to networks with the adversary's route installed. Longest-prefix-match forwarding then causes the traffic to detour to the adversary (see Fig.~\ref{fig:diagram}).\footnote{One potential countermeasure to protect against sub-prefix hijacks is RPKI. However, RPKI deployment is still not ubiquitous~\cite{li2023rovista,nist_rpki_monitor} and flaws in RPKI configurations still leave networks vulnerable to sub-prefix attacks~\cite{gilad2017max_length}.}

\subsection{Adversary Strategy}
Combining these two techniques leads to the following attack strategy:
\begin{enumerate}
    \item Select a victim prefix for the attack.
    \item Use public BGP data to see which major transit providers carry traffic to that prefix (this can be trivially derived by seeing the most commonly used upstream ASes used for that prefix).
    \item Rent a VM at an IXP where several of those transit providers colocate.
    \item Approach these transit providers and request a BGP transit session.
    \item Use social engineering and manipulation of routing data to trick these transit providers into whitelisting a sub-prefix of the victim's prefix.
    \item Announce a malicious sub-prefix of the victim's prefix with the NO\_EXPORT community.

\end{enumerate}

This attack strategy is very effective as it affects traffic from a large portion of the Internet (potentially 100\%, if an adversary installs the malicious route in all of a victim's neighbors) while being completely stealthy. Furthermore, although some steps, like the establishment of BGP transit sessions, bear a cost, this attack's stealth makes this infrastructure significantly more reusable. Presuming the attack goes undetected, the reputation cost of the attack is zero. This allows an adversary to potentially reuse its ASN, IXP VM, and transit links for many attacks. Stealthy attacks can avoid the negative reputation that often causes adversary infrastructure to get destroyed~\cite{bitcanal_nanog_discussion}.

\section{Ethically Launching the Attack}

We ethically launched this attack on the Internet using the cloud provider Vultr. Vultr allows customers with their own IP address space to make BGP announcements. Using space allocated by our affiliated institution, we established BGP sessions with Vultr's routers from virtual machines at various Vultr data centers. We relied on the BIRD BGP daemon v1.6.8 running on Ubuntu 22.04 to handle the sessions with the Vultr routers and make BGP announcements. Also, note that Vultr operates several topologically disjoint data centers and does not use any private backbone for customer traffic. Because of this, these data centers can be thought of as distinct BGP-speaking nodes. With this real-world attack, we confirmed that the NO\_EXPORT community offers stealth from monitoring services while allowing us to install a route into a major transit provider's network.

Our experimentation was completely ethical in that 1) all nodes making BGP announcements for these prefixes (including the ones being used as the adversary) had proper authorization to make these announcements, 2) the IP prefixes used were allocated specifically for research purposes and operated no real network services, 3) our experiment involved very few BGP announcements that did not pose any undue burden on production routers, and 4) we only used standard BGP attributes and community values that are commonly used in production routing configurations.

\vspace{.5cm}
\myitem{Normal Routing:} We began by making a standard BGP announcement from Vultr's Silicon Valley data center for the prefix A.B.C.0/23 to the provider Cogent, which provides transit to that data center. This announcement provided routing for our  webserver that was hosted in the Silicon Valley data center and served as the hypothetical victim's server. We confirmed this announcement had converged using various BGP looking glasses~\cite{ntt_looking_glass,cogent_looking_glass} (services provided by networks that allow operators to query routes used by those networks), and we confirmed data-plane connectivity to the webserver. At this point we configured four BGP monitoring services (Cisco Crosswork Cloud, ThousandEyes, RIPE RIS, and RouteViews) to monitor the victim's announcement and throw an alert if any sub-prefix announcements were noticed. We also configured a testing client in Vultr's São Paulo data center that had a BGP session with Vultr's upstream router that we used to 1) test connectivity to the victim's webserver and 2) use ``show route'' on our BGP router to observe this client's control-plane route to the victim's prefix. As expected, this client correctly routed to the victim's webserver and had installed the proper control-plane route as announced by the victim.

\vspace{.5cm}
\myitem{Non-Stealthy Attack:} To verify our monitoring configuration, we made a non-stealthy control BGP hijack from the Vultr data center in New Jersey (which served as the adversary in our experiment) for a sub-prefix of the victim's prefix (A.B.C.0/24). This attack was effective in that traffic from our test client in São Paulo (as well as any other clients tested) was routed to the adversary. However, all control-plane monitors detected this attack. All three monitoring services generated notifications, and the Bird router at São Paulo had installed the adversary's prefix. Thus, while capable of hijacking the victim, this attack was highly noticeable.

\vspace{.5cm}
\myitem{Stealthy Attack:} We withdrew this control announcement and updated our adversary's configurations to launch a stealthy attack, as discussed above. To do this, we attached BGP communities supported by Vultr to instruct it to only export our route to the provider Cogent. Cogent (AS 174) offers a Cogent-specific version (value 174:990) of the RFC 1997 NO\_EXPORT community as specified in their routing guide for customers~\cite{cogent_communities}. This was crucial for enabling our experiments as we were not a direct customer of Cogent but instead had to propagate our announcements through Vultr first.\footnote{This was because we did not want to invest in obtaining a tier-1 provider just for this experiment. However, the cost of a BGP transit session is minuscule compared to the gains offered by a BGP attack~\cite{birgelee2022klayswap,coinbase2022celer_bridge} making this a justifiable expense for the adversary.} Had we attached the RFC 1997 NO\_EXPORT value to our BGP announcements, Vultr's border routers would interpret this community and apply export restrictions coming out of \emph{Vultr's} network, preventing our announcement from ever reaching Cogent. On the other hand, the Cogent-specific NO\_EXPORT value did not have any meaning to Vultr's routers and was transparently transited to Cogent where the Cogent routers interpreted this community.

The adversary announcement for launching this attack was for the prefix A.B.C.0/24 and included a set of communities supported by Vultr that instructed Vultr only to export the adversary's route to Cogent and the 174:990 community supported by Cogent, preventing the exporting of the route beyond Cogent's network. In this manner, the stealthy attack was only in the route tables of Vultr's NJ data center and Cogent.

\subsection{Attack Measurements}

We confirmed this attack was both stealthy and effective via several measurements. We started by measuring the behavior of our São Paulo client, which routed traffic to the adversary instead of the victim. We used ``show route'' to examine the control plane route head by the client and found \textbf{the router still only showed the victim's announcement, meaning this client was affected by the adversary's attack while not hearing the malicious announcement in the control plane}. We expanded our measurements to look at the four BGP monitoring services we studied and found that none of them had thrown an alarm for the adversary's route. We further confirmed that Cogent was a peer of RIPE RIS (as specified in the RIS peer list~\cite{ris_peers}) and during our control announcement Cogent's routers had exported the adversary route over this direct peering session. However, \textbf{when the NO\_EXPORT community was applied, exporting of the route over the peering session with RIPE RIS was indeed suppressed}.\footnote{Cogent did not have a direct peering session with any of the other BGP monitoring services. Routes for the sub-prefix were suppressed in other monitoring services as well.} We further confirmed that the adversary's route was installed in Cogent routers via the Cogent looking glass~\cite{cogent_looking_glass}\footnote{The looking glass is not a BGP monitoring service as it is intended for manual debugging of routes and requires a high-latency query to be run for each route lookup command. Some looking glasses even explicitly prohibit automated queries~\cite{ntt_looking_glass}. Thus, we do not consider the presence of the sub-prefix route in the looking glass as  attack detection by BGP monitoring.} demonstrating the ability to infect a network with a malicious route and suppress that network's reporting of the route to BGP monitoring services.

Having confirmed this announcement was fully stealthy to all the monitoring services we tested, we wanted to measure how much of the Internet was using the adversary's route unknowingly. To do this we took a sample of 4096 random IP addresses and PING scanned them with the command ``nmap -sn -iR 4096''. We are conscious of the ethical concerns with random scanning and ensured our technique was ethical by not performing any port scanning (preventing any behavior that could be seen as possibly probing for vulnerable services) and sending only a very small number ($\sim$3) of ICMP Echo requests to each host. We felt this scan was unlikely to raise any security alarms even on highly-monitored systems (as any public IP address is likely to be scanned several times as part of many projects) and imposed a negligible computational load even for under-provisioned systems. From this scan, we randomly selected 1000 responsive hosts to use in subsequent scans related to our attack.\footnote{Not all hosts responded in subsequent scans, which reduced the effective sample size for each scan, but for consistency, we always sent to this same 1k host sample.} This sample is representative of IPv4 addresses that respond to ICMP Echo.

With the adversary's attack active, we sent out ICMP Echo requests to the host in our sample from the victim's machine using a source IP address in the hijacked prefix. Thus, the host's ICMP Echo responses would be routed to the victim or the adversary, depending on whether that host's network was affected by the attack. When scanning the 1k sample, 739 hosts responded, all of which routed to the adversary (i.e., the victim was listening for responses but did not hear any) \textbf{indicating that the attack had global affect despite it being invisible to the route monitoring services we studied}.

\section{Broader Viability of the Attack}

\subsection{NO\_EXPORT Behavior at Tier-1 ISPs}
\label{sec:network_support}
We studied the NO\_EXPORT community behavior required to launch this attack at several tier-1 ISPs and confirmed the vulnerable behavior at all networks we studied. Even though RFC 1997 states that any BGP speaker that understands communities shall implement well-known communities,  networks could conceivably strip these communities from BGP route announcements, or filter routes with these communities. Checking for these practices would require a direct connection to the associated network---not possible with our Vultr test environment.
Instead, we reached out to the operators of several other tier-2 providers to assist us by making BGP announcements tagged with the NO\_EXPORT community to their tier-1 providers. Through this technique, we were able to confirm the behavior of the community at Arelion (AS 1299), NTT (AS 2914), and Sprintlink (AS 1239), in addition to Cogent (AS 174, which we tested through Vultr). At these networks, we confirmed the following behavior:

\begin{itemize}
    \item These networks supported the NO\_EXPORT community and installed routes with this community into their route tables (verified via looking glasses).
    \item These networks had a direct peering session with RIPE RIS~\cite{ris_peers} and RouteViews~\cite{routeviews_peers}, except for Cogent, which only peered with RIPE RIS.
    \item These networks properly exported routes not tagged with the NO\_EXPORT community to BGP monitoring but did not export our test announcements that contained the NO\_EXPORT community.
    \item The announcements through these networks were \emph{not} seen in Crosswork Cloud or ThousandEyes.
\end{itemize}

The networks we tested were all tier-1 ISPs that are on a significant number of paths. Even though it is difficult to scale these tests to large numbers of networks,
our results strongly suggest widespread viability of this attack. Based on the many networks known to support BGP communities~\cite{bgp-comm-support}, the language in RFC 1997 (``[these communities] operations shall be implemented in any community-attribute-aware BGP speaker''), and instructions for peering from BGP monitoring services~\cite{ris_peering_policy, thousandeyes_private_bgp, routeviews_peering}, we believe the vast majority of networks are likely susceptible to this attack.

\subsection{Effective Spread of the Attack Using Simulations}

We ran Internet topology simulations using the Gao-Rexford model of routing policies~\cite{gao2001gao_rexford} to study the AS-level paths of Internet traffic between 150 randomly-chosen ASes. We measured the faction of paths that contained an AS with the adversary's route installed under various different attack scenarios. We found that
\textbf{by making its announcement to only five tier-1 ISPs, an adversary can on average hijack 37\% of Internet traffic to a victim destination}. Thus an adversary can still affect a large portion of the Internet even with the NO\_EXPORT community limiting the spread of its announcement. Additionally, if the adversary announced to only the networks at which we confirmed the behavior of the NO\_EXPORT community, the adversary could still on average hijack traffic from 23\% of the Internet. This vastly outperforms the capabilities of previous stealthy attacks (which often affected less than 2\% of the Internet). We further discusses our methodology and results in Appendix~\ref{app:simulations}.

\section{Defenses}

\subsection{Changes in BGP Configuration}

At its core, this attack revolves around the fact that the well-known NO\_EXPORT community restricts exporting to BGP monitoring sessions. The most immediate mitigation to this attack is to change that behavior. This behavior can be changed without even restricting the use of the NO\_EXPORT community or changing the way BGP monitoring is configured. Instead, the RFC 1997 value for the NO\_EXPORT community can be rewritten (via an ingress policy on eBGP sessions with external ASes) to a different AS-specific community. This community can then be matched on egress sessions with customers, peers, and providers to restrict exporting. On BGP sessions with monitoring services however, this community would not have any effect. We discuss additional configuration options to mitigate these attacks and show configuration examples in Appendix~\ref{app:configs}.

\subsection{Use of BGP Monitoring Protocol (BMP)}
The BGP Monitoring Protocol (BMP)~\cite{bmp_rfc} is a protocol specifically designed for the monitoring of BGP updates and RIBs that can expose prefixes tagged with NO\_EXPORT. BMP could potentially be run between peer networks and monitoring services in place of the standard eBGP multi-hop sessions thus allowing for full visibility of the RIBs of participating networks by monitoring services. BMP is supported in many leading BGP implementations~\cite{cisco_bmp,juniper_bmp} and has stand-alone implementations as well~\cite{open_bmp}.

\subsection{Increased Scope of BGP Monitoring}

While the changes in BGP configurations and monitoring protocols proposed above vastly reduce the threat surface of this attack (as many major networks peer with BGP monitoring services), there is still a chance for an adversary to launch the attack at networks that do not peer with BGP monitoring services. Specifically, imagine the path between a traffic source and a victim AS pass through a transit provider $T$ that does not peer with any BGP monitoring services. An adversary could still use the NO\_EXPORT community to infect $T$ with a malicious route and since $T$ is not peering with a BGP monitoring service, the attack would remain stealthy. To this end, it is important that BGP monitoring services continue to expand the number of peers they use and attempt to peer with as many transit ASes as possible.

\subsection{Use of RPKI}
RPKI offers a cryptographic database of IP prefix ownership~\cite{RFC6480} which can be used to filter BGP announcements~\cite{RFC_rov}. In addition to defining prefix ownership, RPKI defines the specific lengths of the IP prefixes that can be announced in BGP~\cite{gilad2017max_length}. This can potentially prevent our stealthy BGP attack which is based on a sub-prefix attack (although misuse of the max-length attribute can degrade this protection~\cite{gilad2017max_length}). This protection holds even if an adversary bypasses traditional prefix ownership checks by prepending the victim's ASN to its announcement~\cite{dfoh}. Thus, proper RPKI use is a strong mitigation for stealthy BGP attacks.

\section{Related Work}
\subsection{Stealthy BGP Attacks}
Milolidakis \emph{et al.} studied smart BGP attacks that evaded public route collectors but used a vastly different mechanism to launch their attack (selective announcement to different BGP neighbors) and did not consider BGP communities or utilize any mechanism to prevent ASes that were peering with monitoring services from exporting the malicious route~\cite{milolidakis2021smart_hijacks}. As these attacks must be limited only to ASes not peering with monitoring services, these attacks can only affect roughly 2\% of the Internet in favorable circumstances.

Morillo \emph{et al.} discuss hidden BGP hijacks that are not visible to networks affected by the hijacks because upstream networks perform ROV filtering on the malicious announcements~\cite{morillo2021rov++}. Our attack similarly affects networks that do not have the malicious route installed, but in our attacks, the adversary intentionally limits the spread of the announcement via the NO\_EXPORT community. Furthermore, while hidden hijacks are hidden from some ASes using the malicious route, there is no guarantee that they are hidden from monitoring services. Hidden hijacks involve the adversary announcing a standard, unrestricted, sub-prefix hijack (which usually has near global visibility in BGP monitoring).

Heng discussed a BGP attack on a real network that evaded monitoring services by being localized to only a single source network (Yahoo mail)~\cite{invisible_hijack}. While this attack did affect a strategic target of interest to the adversary, it did not affect a significant portion of the Internet. Further, it can be seen as a real-world example of a smart BGP attack as discussed by Milolidakis \emph{et al.}. The malicious route used in the attack was not installed by any networks that were peering with monitoring services, and the adversary did not use any communities to control the propagation of its announcement. By use of the NO\_EXPORT community, our attack has the potential to affect significantly more networks and remain stealthy.

Birge-Lee \emph{et al.} discuss using BGP communities on equally-specific attacks to both enable BGP interception attacks and limit the spread of BGP attacks~\cite{birgelee2019sico}. Our work has vastly different means and objectives than the previous work by Birge-Lee \emph{et al.}. The stealthy attacks presented by Birge-Lee \emph{et al.} aimed to affect a single target of interest and \textbf{as few other hosts on the Internet as possible}, and this was realized by using BGP communities to shape equally-specific BGP attacks. Furthermore, the ability of these attacks to specifically evade BGP monitoring systems was not studied. Our work has a completely opposite objective (i.e., to affect as much of the Internet as possible, not just a single host) and achieves this using a sub-prefix BGP attack.

Miller \emph{et al.} use the NO\_EXPORT community in the context of BGP attacks that exploit traffic blackholing~\cite{miller2019blackhole}. They do mention the community's ability to limit announcement spread and the use of the NO\_EXPORT community on sub-prefix BGP attack in conjunction with blackhole communities. However, they do not study the effect on BGP monitoring systems and, most crucially, do not come to the key insight that even ASes which provide data to BGP monitoring systems will inappropriately suppress information on the adversary's attack when the NO\_EXPORT community is attached. Furthermore, using the NO\_EXPORT community on blackhole announcements (where it is already a recommended best practice~\cite{miller2019blackhole}) is significantly different than the use of the NO\_EXPORT community on BGP hijacking announcements that actually maliciously deliver traffic to the adversary.

\subsection{BGP Detection Systems}
Given data from BGP monitoring, several works propose algorithms for detecting BGP attacks~\cite{dfoh,phas_bgp_monitoring,sermpezis2018artemis}. Our attacks' ability to achieve stealth is \emph{independent} of what detection algorithm is used for processing BGP data as \emph{none} of the adversary's malicious BGP announcements are seen by monitoring services. We do not exploit any specific BGP attack detection algorithm behavior but instead achieve stealth by preventing the adversary's malicious announcements from appearing in the BGP data fed into these algorithms in the first place.

\section{Conclusion}

This paper challenges the assumption that stealthy BGP hijacks are not globally effective and highlights the need to harden BGP monitoring based on our recommendations. Stealthy BGP attacks can be launched by using the well-known NO\_EXPORT community on a sub-prefix BGP hijack. These attacks are are highly effective at hijacking vast amounts of Internet traffic while remaining invisible to leading BGP monitoring services. These attacks are also not seen by affected ASes that are forwarding traffic to the adversary. While this is due to the RFC-specified behavior of the well-known NO\_EXPORT community, simple changes to router configurations can mitigate this attack. These changes cause ASes peering with BGP monitoring services to still report routes tagged with the NO\_EXPORT community. However, even with these changes it is essential that BGP monitoring services peer with as many transit providers as possible. More broadly, the push for fundamentally more secure interdomain routing is essential as BGP monitoring will always be limited by the number of monitors available and what routes those monitors see. However, in the meantime as we wait for robust solutions, securing ASes against stealthy BGP hijacks  is essential.

\bibliographystyle{ACM-Reference-Format}
\bibliography{refs}

\appendix

\section{Ethics}

We took care to ensure that all of research was ethical. All of our simulation measurements used publicly-accessible data and did not derive any conclusions about individual users (that could be considered a privacy violation). Our ethical real-world attack was achieved by \textbf{attacking ourselves} and using only nodes (including the nodes we used for the adversary) that were authorized to announce our IP prefix. We obtained an IP prefix specifically set aside for network research from our affiliated institution (that ran no real network services). Authorization for the use and BGP announcements of this prefix was given in a Letter of Authorization (LoA) signed by the relevant office at our institution. This LoA as well as the Internet Routing Registry (IRR) entries for this prefix contained all origin ASes used to originate the prefix in our experiments and, based on the LoA and the IRR entries, all of our upstream allowed our announcements. We took care to only make announcements at a reasonable rate. Furthermore, given the well-known NO\_EXPORT community is RFC standardized and SHOULD be supported by all BGP speakers that understand communities, we did not see any risk in sending out a BGP announcement with this community attached. We also conducted random Internet scanning in this paper, but we made sure this scanning did not pose a volume, security, or privacy threat to the networks being scanned or the providers we used for the scanning. We scanned with a small sample (4096 hosts initially that was reduced to 1k hosts based on responses) using an extremely small number of packets (roughly 3 ICMP Echo packets) that were sent at a reasonable rate. This small frequency of packets does not pose an undue load even on dated or under-provisioned equipment. We did not do any port scanning or scanning for support of protocols besides ICMP Echo which prevented our scan from looking like an intrusion attempt. Finally, we did not record any information from the hosts that responded other than if they responded and if they sent to the victim or adversary during our attack.

\section{Simulations}
\label{app:simulations}
To understand how much of the Internet this adversary could stealthily hijack, we then ran Internet topology simulations using the Gao-Rexford model of routing policies~\cite{gao2001gao_rexford} on the CAIDA AS-Relationships Data Set~\cite{caida}. We simulated all routes between 150 randomly chosen ASes from the CAIDA topology. For each of the AS-level routes generated, we tested to see if one of the ASes the adversary could compromise with a stealthy attack was in the simulated path. If we so, we considered this path compromised as the source of the traffic would unknowingly send its traffic to an AS that had a stealthy attack installed in its route table (which would forward the traffic to the adversary). We grouped the data by origin AS allowing us to obtain a percentage of traffic sources (i.e., source ASes) affected by the stealthy hijack for each origin AS.

\begin{figure}[t]
    \centering
    \includegraphics[width=0.9\linewidth]{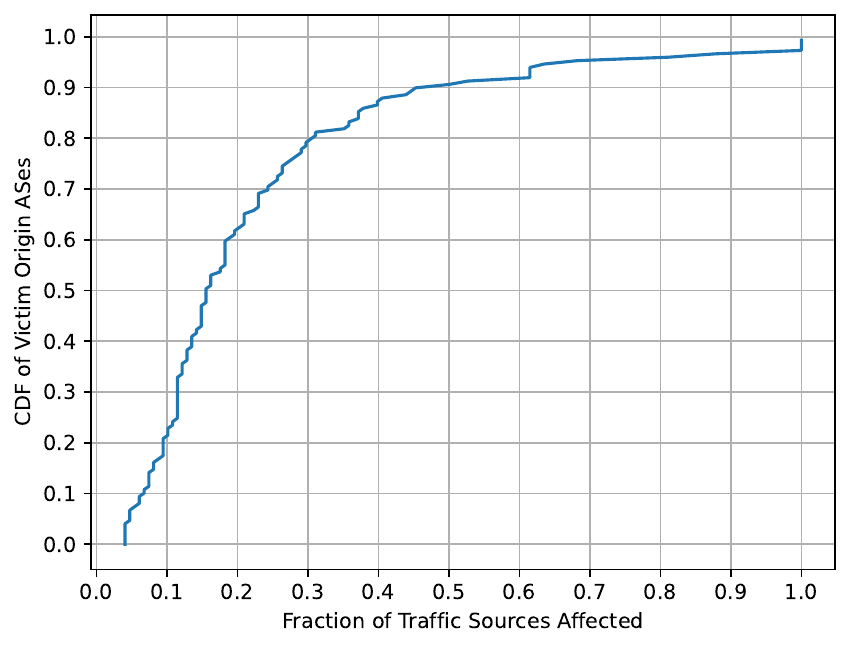}
    \caption{A CDF of the fraction of traffic sources affected by a stealthy hijack launched by an adversary that installed malicious routes at the four ASes tested in this paper (that all are found to be vulnerable to the stealthy hijack).}
    \label{fig:known-ases}
    \vspace{5pt}
\end{figure}

We began by considering an adversary that only installed its route at the four networks we found vulnerable to the attack in Sec.~\ref{sec:network_support}: AS 1299, AS 2914, AS 1239, and AS 174. \textbf{Using only these networks, which we confirmed are vulnerable, an adversary could, on average, hijack traffic from 23\% of the Internet.} The CDF of the percent of Internet traffic that can be hijacked by this adversary when targeting different victim destination ASes is shown in Fig.~\ref{fig:known-ases}. Notably, some victim ASes (represented at the top right of the CDF) that used exclusively compromised providers for Internet connectivity have 100\% of their traffic affected by the stealthy attack, representing the global effectiveness with 0\% observability example we saw in our ethical real-world attack.

\begin{figure}[t]
    \centering
    \includegraphics[width=0.9\linewidth]{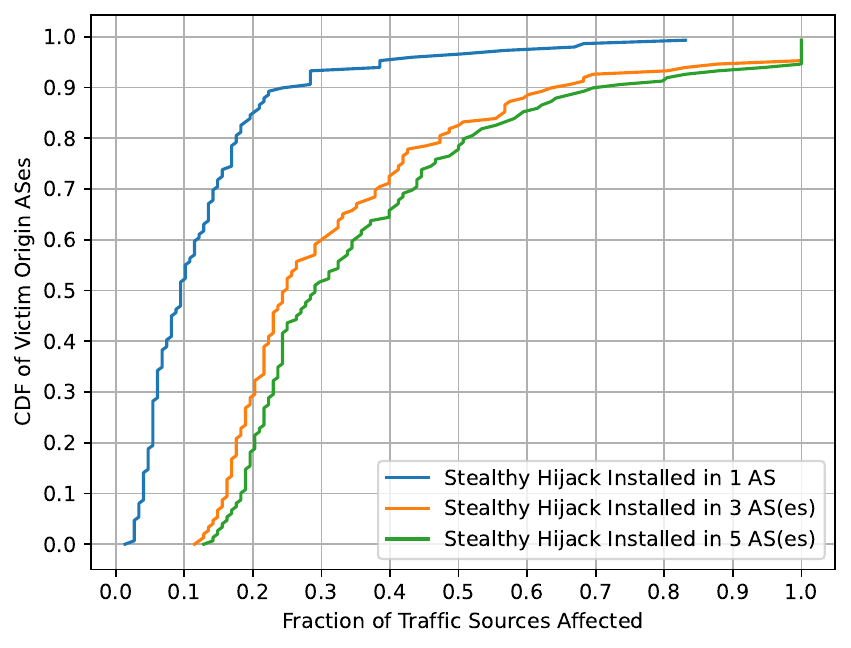}
    \caption{A CDF of the fraction of traffic sources affected by a stealthy hijack launched by an adversary that installed malicious routes at the top 1, 3, and 5 networks by customer cone size.}
    \label{fig:top-ases}
    \vspace{5pt}
\end{figure}

Given the behavior we exploit is stipulated in RFC 1997 (so all ASes capable of understanding communities should exhibit this behavior) and all ASes studied correctly conformed to this RFC, it is reasonable to assume the vast majority of major networks behave in this manner, particularly given the large amount of BGP community support~\cite{bgp-comm-support,birgelee2019sico}. Thus, we ran our simulation considering a strategic adversary that chooses ASes based on the size of their customer cones (i.e., the number of direct and indirect customers that AS has). Based on the CAIDA AS-rank customer cone counts~\cite{caida_as_rank}, we modeled an adversary that compromised a varying number of the top ASes (in order of customer-cone size: AS 3356 Lumen, AS 1299 Arelion, AS 174 Cogent, AS 2914 NTT, AS 6762 Telecom Italia Sparkle). If an adversary only installs its stealthy hijack at the largest AS: 3356 (Lumen), it, on average, can affect routes to 0.13\% of destinations. Using only the top 2 networks, an adversary can affect 18\% of destinations on average. This amount increases to 34\%, 35\% and 37\% if the adversary installs its route in the top 3, 4, or 5 networks, respectively. A CDF of the effectiveness of these hijacks is shown in Fig.~\ref{fig:top-ases}.

Overall, these simulations confirm that even if an adversary can only install its attack at a limited number of top networks, significant swaths of Internet traffic can be hijacked in a stealthy manner making this hijack significantly more viable than previous techniques for launching stealthy BGP hijacks~\cite{milolidakis2021smart_hijacks}.

\section{Config Examples}
\label{app:configs}
This is an example of pseudocode for a router configuration that rewrites the NO\_EXPORT community at ingress to an AS-specific community that matches the original NO\_EXPORT behavior (except in the case of sessions with BGP monitoring services).

\begin{verbatim}
filter neighbor_in {
    if NO_EXPORT in bgp_communities {
        bgp_communities.remove(NO_EXPORT);
        bgp_communities.add(ASN:123);
    }
    # Similar blocks need to also be include
    # for NO_ADVERTISE and NO_EXPORT_SUBCONFED
    ...
    accept;
}

filter neighbor_out {
    if ASN:123 in bgp_communities {
        reject;
    }
    accept;
}

protocol bgp neighbor {
    ...
    import filter neighbor_in;
    export filter neighbor_out;
}

protocol bgp  bgp_monitoring_service {
    ...
    import none;
    export all;
}

\end{verbatim}

Another potential configuration change to mitigate this attack is to run BGP sessions with monitoring services as Internal BGP (iBGP) sessions instead of eBGP sessions. This way, routers would no longer consider BGP sessions with monitoring services as crossing an AS boundary, and thus, the export restrictions of the NO\_EXPORT community would not apply. There still may be some attack surface left open by the RFC-standardized NO\_ADVERTISE community~\cite{RFC1997} (which limits exporting to any other routers even in the same AS), but these attacks are likely significantly less effective. If an adversary were to use the NO\_ADVERTISE community to launch an attack its malicious announcement would be localized to only a single router.

\end{document}